# The effect of substitution of Mn for Ni on the magnetic and transport properties of CeNi$_{0.8-x}$Mn$_x$Bi$_2$

Soo-Whan Kim[1], Kyu Joon Lee[1], D.T. Adroja[2,3], F. Demmel[2], J.W. Taylor[2] and Myung-Hwa Jung[1*]

[1] Department of Physics, Sogang University, Seoul 121-742, South Korea

[2] ISIS Facility, Rutherford Appleton Laboratory, Chilton, Didcot Oxon OX11 0QX, UK

[3] Physics Department, University of Johannesburg, PO Box 524 Auckland Park 2006, South Africa

*Corresponding author: E-mail: mhjung@sogang.ac.kr

**Abstract**

We report the results of Mn substitution for Ni in CeNi$_{0.8}$Bi$_2$, (i.e. CeNi$_{0.8-x}$Mn$_x$Bi$_2$). All the samples have an antiferromagnetic ordered state below $T_N$ = 5.0 K due to localized 4f-magnetic moment on the Ce ions. Besides this antiferromagnetic ordering caused by Ce, the magnetic and transport properties are abruptly changed with increasing Mn contents at the boundary composition of $x$ = 0.4. The magnetic state is changed into a ferromagnetic state around 200 K for $x > 0.4$, where the electrical resistivity is strongly suppressed to become simple metallic. These results of ferromagnetism and metallicity can be explained by the double-exchange mechanism. The mixed valence states of Ni and Mn ions are confirmed by X-ray photoelectron spectroscopy (XPS). For $x \leq 0.4$, the initial Ni$^{3+}$ state gradually changes to the Ni$^{2+}$ state with increasing $x$ up to 0.4. On further increase of $x > 0.4$, the Ni$^{2+}$ state is substituted for the Mn$^{2+}$ state, which gradually changes to the final Mn$^{3+}$ state. We also present an inelastic neutron scattering (INS) measurements on CeNi$_{0.8}$Bi$_2$ (i.e. x=0) between 1.2 and 12 K. The high energy INS study reveals the presence of two well defined crystal electric field (CEF) excitations near 9 meV and 19 meV at 1.2 K and 6 K, while the low energy INS study reveals the presence of quasi-elastic scattering above 4 K. We will discuss our INS results of CeNi$_{0.8}$Bi$_2$ based on the crystal electric field model.

# 1. Introduction

Ce-based compounds have been the concern of continuous interests because of the exotic physical properties such as Kondo effect [1], various magnetic states [2, 3, 4], mixed valence or valence fluctuations behavior [5], heavy fermion [6], unconventional superconductivity [7], and non-Fermi liquid behavior [8]. These properties are mainly controlled by two competing order parameters, long range intersite Ruderman-Kittel-Kasuya-Yosida (RKKY temperature, $T_{RKKY}$) interactions and screening of 4f local magnetic moments through onsite Kondo interactions (Kondo temperature, $T_K$). The competing strength is governed by the exchange interaction $J_{cf}$ between localized 4f magnetic moments and itinerant conduction electrons. The competition between $T_{RKKY}$ and $T_K$ versus $J_{cf}$ is qualitatively described by the Doniach's phase diagram [9], where $T_{RKKY}$ is proportional to $J_{cf}^2$ and $T_K$ exhibits an exponential dependence of $J_{cf}$. The change of $J_{cf}$ can be achieved simply by applying external pressure or introducing chemical dopants. For example, $CeNiGe_2$ was carefully studied to be an antiferromagnetic Kondo lattice compound [10]. By substituting Co for Ni, the antiferromagnetic ordering is suppressed and the Fermi liquid state is generated. This substitution effect is well explained by the Doniach phase diagram, where the RKKY interaction ($T_{RKKY}$) becomes weaker and the Kondo interaction ($T_K$) becomes stronger as the exchange interaction ($J_{cf}$) is increased with the Co contents. These observations imply that the role of transition metal in the Ce-based compounds is important because most of physical properties are governed by the localized Ce moments interacting with itinerant conduction electrons.

Another Ce-based compound, stoichiometric $CeNiBi_2$, was reported to be also an antiferromagnetic Kondo lattice compound. It has an antiferromagnetic ordering with the Neel temperature of $T_N = 6.0$ K due to the localized Ce moments. The electrical resistivity shows metallic behavior with a broad shoulder resulting from the Kondo and crystal electric field (CEF) effect [11]. On the other hand, Ni-deficient compound of $CeNi_{0.8}Bi_2$ exhibits a superconducting behavior below $T_{sc} = 4.2$ K, in addition to the antiferromagnetic ordering at $T_N = 5.0$ K [12]. Since the superconducting transition temperature $T_{sc}$ is rather higher, compared with the other heavy-fermion superconductors, it has attracted much attention. Later, this superconductivity was alleged to be not intrinsic. The source of the superconductivity is supposed to be Ni deficiency or secondary phase of $NiBi_3$ [13, 14]. Now, there is a question about the transition-metal effect on those physical properties in the $CeNi_{0.8}Bi_2$ compound. In this study, we present the substitution effect of Mn for Ni in a form of $CeNi_{0.8-x}Mn_xBi_2$. This new class of compounds may give the suppression of incomplete superconductivity and can bring new physical properties. We clearly observe that the magnetic and transport properties are abruptly changed around x = 0.4. The magnetization is strongly enhanced with increasing x, and thereby ferromagnetic behavior is observed at x ≥ 0.4. The electrical resistivity is strongly reduced about two orders of magnitude from x = 0 to x = 0.8. In addition, the magnetoresistance is changed from a simple positive curve to a complex negative curve at x = 0.4. These observations can be understood by the mixed valence state of transition metals of Ni and Mn, based on the experimental results of X-ray photoelectron spectroscopy. Furthermore, we study the CEF effect by measuring INS to confirm the magnetic system of $CeNi_{0.8}Bi_2$ at low temperature.

## 2. Methods

Polycrystalline samples of $CeNi_{0.8-x}Mn_xBi_2$ were prepared by solid-state reaction in an evacuated quartz tube. Stoichiometric mixture of starting elements was slowly heated up to 773 K and kept for 10 h, followed by heat treatment at 1023 K for 20 h and at 1073 K for 20 h, which is similar to the method described in Ref. 2. The obtained mixture was well ground and pressed into pellets. The pellets were heated again at 1073 K for 10 h. In order to prevent the oxidation, all the preparation procedures were performed under argon gas environment in a grove box. This procedure was repeated until we obtain homogeneous phase by checking the X-ray diffraction patterns. X-ray powder diffraction with the use of a Rigaku Miniflex II diffractometer using Cu-Kα radiation at 30 kV and 15 mA was applied to characterize the structure type and the phase identification. The diffraction patterns showed that the structure of these samples is tetragonal, including a small amount of impurities such as Bi and $NiBi_3$. The substitution effect on the lattice parameters is negligible because the shift of the peak is not observed with increasing x. The magnetic properties were investigated using a superconducting quantum interference device-vibration sample magnetometer (SQUID-VSM) in the temperature range from 2 to 300 K and applied magnetic fields up to 70 kOe. The transport properties were measured using a standard four-probe method. The valence states of the samples were studied by X-ray photoelectron spectroscopy (XPS) using Thermo K-Alpha. Inelastic neutron scattering (INS) measurements on $CeNi_{0.8}Bi_2$ and $LaNi_{0.8}Bi_2$ were carried out using the time of flight (TOF) chopper spectrometer MARI (direct geometry) and the TOF crystal analyser spectrometer OSIRIS (indirect geometry) between 1.2 K and 12 K. On MARI we use neutrons with incident energy $E_i$ = 15 and 50 meV, while on OSIRIS we used PG002 analyzer with fixed final energy $E_f$ = 1.84 meV.

## 3. Results and Discussion

Figure 1 shows the temperature dependence of magnetization M(T) of $CeNi_{0.8-x}Mn_xBi_2$ (x = 0, 0.1, 0.4, 0.5, 0.7 and 0.8) between 2 K and 300 K in an applied field of 1 kOe. The absolute value of M(T) slowly increases with increasing x up to x = 0.4, and it is abruptly enhanced after x = 0.4. There is a peak structure around $T_N$ = 5.0 K with a kink around 7.2 K which is the onset temperature of the antiferromagnetic order. These antiferromagnetic features persist for all the samples from x = 0 to 0.8. In addition to the antiferromagnetic transition at $T_N$ = 5.0 K, for x = 0, there is an additional transition of superconductivity with $T_{sc}$ = 4.2 K, which is shown in the inset of figure 1. This superconducting feature of $CeNi_{0.8}Bi_2$ turns out to be not intrinsic, but comes from impurity phase such as $NiBi_3$ because the transition is not reproducible and the transition temperature is similar to $T_{sc}$ of the impurity [14]. In our case, we calculate the superconducting volume fraction of about 20% for the x = 0 sample. This indirectly implies that the superconductivity comes from the impurity phase of $NiBi_3$, rather than the intrinsic nature of the main phase of $CeNi_{0.8}Bi_2$. In this sense, we neglect the data related with the superconductivity here from. For x = 0.1 and 0.4, the magnetization slightly increases as x increases. Above 100 K, the magnetization data can be fitted by the Curie-Weiss law, giving rise to the effective magnetic moment of $\mu_{eff}$ = 2.55 $\mu_B$ for all the samples with x ≤ 0.4. This value agrees with 2.54 $\mu_B$ calculated for a free $Ce^{3+}$ ion, indicating that the magnetic moment comes from the trivalent Ce ion state. The obtained paramagnetic Curie temperature decreases from $\theta_P$ = -43 K for x = 0 to -28 K for x = 0.1 and -23 K for x = 0.4, suggesting that the antiferromagnetic exchange strength is suppressed with increasing x. For x > 0.4, the magnetization is strongly enhanced. The absolute values are almost two orders of magnitude larger than those for x ≤ 0.4. Especially, the magnetization undergoes rapid increment below 200 K. We mark such increasing spot, where the magnetization rapidly increases, as ferromagnetic transition temperature $T_c$ = 188 K for x = 0.5, 202 K for x = 0.7, and 200 K for x = 0.8. These $T_c$ values are taken from the intersection point of two linear lines in the low and high temperature ranges (see figure 1(a)).

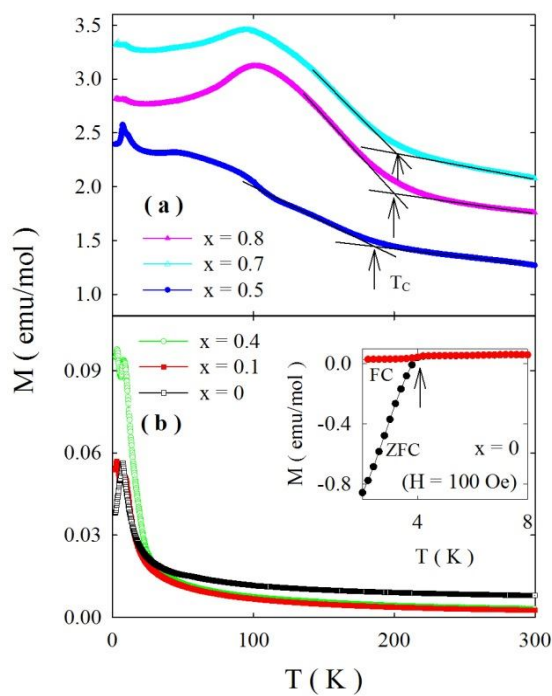

Figure 1. The temperature dependence of magnetization measured at 1 kOe in field-cooled condition for $CeNi_{0.8-x}Mn_xBi_2$ ((a) x = 0.5, 0.7, and 0.8, and (b) x = 0, 0.1, and 0.4). The arrows indicate the ferromagnetic transition temperature derived from two linear lines in the low and high temperature ranges. Inset shows the temperature dependence of magnetization for x = 0, where the arrow represents the superconducting transition temperature.

In order to confirm the ferromagnetic order, we measured the field dependence of magnetization M(H) of $CeNi_{0.8-x}Mn_xBi_2$ at 2 K. In figure 2, the magnetization for x ≤ 0.4 gradually increases with applying field, while the magnetization for x > 0.4 abruptly increases and then nearly saturates above the applied field of 20 kOe. The gradual increment of magnetization for x ≤ 0.4 is attributed to the antiferromagnetic order of preassembly Ce moments. On the other hand, the abrupt increment of magnetization at low fields for x > 0.4 confirms the ferromagnetic order mainly from Mn moments, which need to be confirmed through neutron diffraction measurements. Moreover, the magnetic hysteresis is observed in x > 0.4. One example of x = 0.8 is shown in the inset of figure 2. The clear sign of hysteresis is another evidence of ferromagnetic order in x > 0.4.

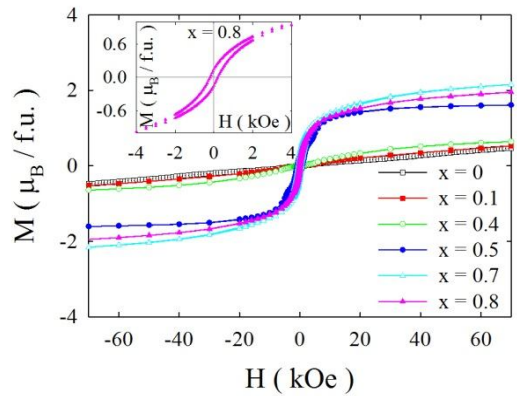

Figure 2. The magnetic field dependence of magnetization measured at 2 K for $CeNi_{0.8-x}Mn_xBi_2$ (x = 0, 0.1, 0.4, 0.5, 0.7 and 0.8). Inset shows the typical magnetic hysteresis loop for x = 0.8.

Since the magnetic properties are abruptly changed at the boundary composition of x = 0.4, we have also studied the changes of transport properties with x. Figure 3 presents the temperature dependence of electrical resistivity ρ(T) of $CeNi_{0.8-x}Mn_xBi_2$. For x = 0, the resistivity exhibits metallic behavior with a broad shoulder at 100 K, resulting from the interplay between Kondo and CEF effects [11, 15].

At low temperatures, the rapid drop at 4.2 K is related with the superconducting transitions from $NiBi_3$ impurity, which was already mentioned in the magnetization data. For x = 0.1, the resistivity curve shows a broad hump around 80 K, which is moved toward lower temperature to 60 K at x = 0.4. This suggests that the CEF effect is suppressed by the Mn contents. The superconducting signal is still observable in x = 0.1, but disappears for x ≥ 0.4. It is noteworthy that as x increases, the absolute value of resistivity tends to be reduced except x = 0.4. For x = 0.5, the resistivity value substantially decreases compared to the x = 0.4 compound. Furthermore, the slope changes near 200 K could be associated with the ferromagnetic ordering as observed through the magnetic susceptibility measurements as shown in figure 1(a). For further increase of x = 0.7 and 0.8, the resistivity shows a simple metallic behavior in all the temperature range from 2 K and 300 K.

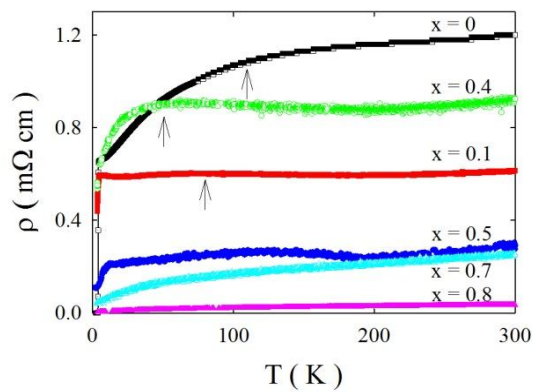

Figure 3. The temperature dependence of electrical resistivity for $CeNi_{0.8-x}Mn_xBi_2$ (x = 0, 0.1, 0.4, 0.5, 0.7 and 0.8). The arrows indicate the position of broad hump related with CEF effect.

In order to obtain more information on the evolution of transport properties by Mn substitution, the magnetoresistance was measured at two different temperatures of 6 K and 300 K. The results are plotted in figure 4(a) for 6 K and (b) for 300 K. The low-temperature data were taken at 6 K in order to rule out the effect of superconducting behavior. Compared with the positive magnetoresistance behavior at 300 K (see figure 4(b)), the magnetoresistance at 6 K exhibits complicated field dependence. The positive MR at 300 K is simply understood by ordinary magnetoresistance behavior.

At 6 K, the samples of x < 0.4 are simply antiferromagnetic, while the samples of x ≥ 0.4 are ferromagnetic. The positive MR behavior for x < 0.4 is attributed to the increase of spin-disorder scattering, because the antiparallel spins tend to be tilted by applying magnetic field. This disorder of spin alignment in the antiferromagnetic state can induce the positive magnetoresistance [16]. For x ≥ 0.4, the positive magnetoresistance changes into negative at high magnetic fields above 20 kOe, where the magnetization starts to saturate. The origin of positive magnetoresistance at low fields can be the spin-disorder scattering. On the other hand, the negative magnetoresistance behavior in the high field range can be related with the ferromagnetic ordering state in x ≥ 0.4. As described in the magnetization data, the magnetic state for x ≥ 0.4 is ferromagnetic. The negative slope of magnetoresistance can be understood by the fact that the magnetic scattering in the parallel spin configuration is weaker when the spins are aligned in the same direction of applied field. This weaker magnetic scattering can induce the negative magnetoresistance. Here, it should be mentioned that the magnetoresistance ratio measured at 300 K is maximum for x = 0.4, which is the boundary composition from antiferromagnetic to ferromagnetic behaviors.

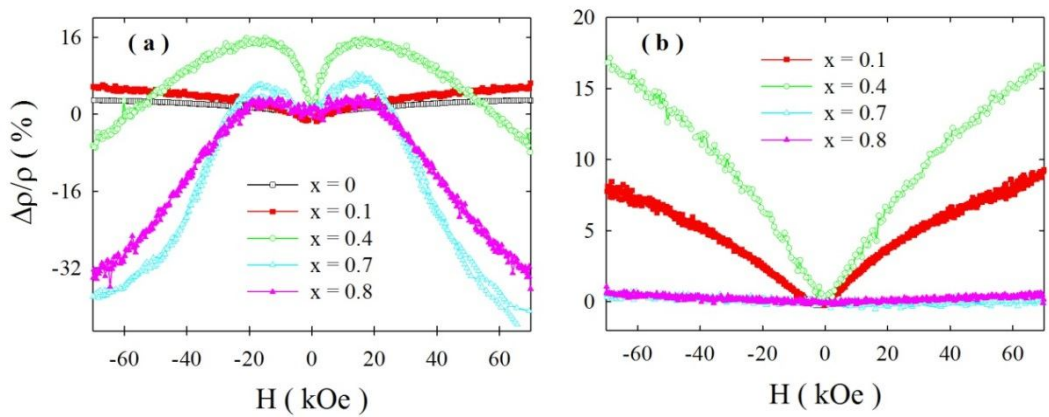

Figure 4. The magnetoresistance curves measured at (a) 6 K and (b) 300 K for CeNi$_{0.8-x}$Mn$_x$Bi$_2$ (x = 0, 0.1, 0.4, 0.7 and 0.8).

Now, let us discuss the origin of such abrupt changes of physical properties at the boundary composition of x = 0.4. One possible scenario to explain the ferromagnetism and metallicity induced by the Mn substitution is double-exchange interaction, which arises between magnetic ions with different valence state. One example is given by the interaction of Mn-O-Mn, in which the $Mn^{3+}$ state interacts with the $Mn^{4+}$ state via oxygen ion. The mixed states of Mn play an important role in ferromagnetic transition accompanied with metallic transition. One of the Mn ions has one more electron than the other, then the electron freely moves from one to the other without changing the spin direction. There are more examples to show ferromagnetism and metallicity via double-exchange mechanism [17, 18], which is more easily achieved by chemical doping. With this scenario, we can conceptually conjecture the double-exchange interaction in our $CeNi_{0.8-x}Mn_xBi_2$ system, where both ferromagnetism and metallicity are induced by the Mn substitution. Thus, we have investigated the valence states of Ni and Mn in $CeNi_{0.8-x}Mn_xBi_2$ by measuring the Mn $2p_{3/2}$ and Ni $2p_{3/2}$ core level XPS spectra. Figure 5 shows the XPS spectra of the Ni $2p_{3/2}$ and Mn $2p_{3/2}$ core levels for x ≤ 0.4 and x ≥ 0.4. In high-energy XPS spectra for x ≤ 0.4 shown in figure 5(a), the peak at 855.8 eV for x = 0.1 corresponds to the $Ni^{3+}(2p_{3/2})$ level, and it slowly shifts to lower energy with increasing x. For x = 0.4, the peak is observed at 854.9 eV, which corresponds to the $Ni^{2+}(2p_{3/2})$ level [19, 20]. These results suggest that the valence state of Ni ion slowly changes from 3+ to 2+ when x increases. On the other hand, the changes of the valence values of Mn ions can be clearly seen in low-energy XPS spectra for x ≥ 0.4, which is shown in figure 5(b). The peak at 640.8 eV for x = 0.4 corresponds to the $Mn^{2+}(2p_{3/2})$ level, and it slowly shifts to higher energy with increasing x. For x = 0.8, the peak is observed at 641.6 eV, which corresponds to the $Mn^{3+}(2p_{3/2})$ level [21, 22]. These results suggest that the valence state of Mn ion slowly changes from 2+ to 3+ when x increases. As a result of figure 5, the initial $Ni^{3+}(3d^7)$ state gradually changes to the $Ni^{2+}(3d^8)$ state with increasing x. On further increase of x, the $Ni^{2+}(3d^8)$ state is substituted for the $Mn^{2+}(3d^5)$ state, which gradually changes to the $Mn^{3+}(3d^4)$ state. These effects could derive the reduction of electrical resistivity by changing the conduction electron concentration and the ferromagnetic ordering because of the double exchange interaction between the

$Mn^{2+}$ and $Mn^{3+}$ states for x > 0.4.

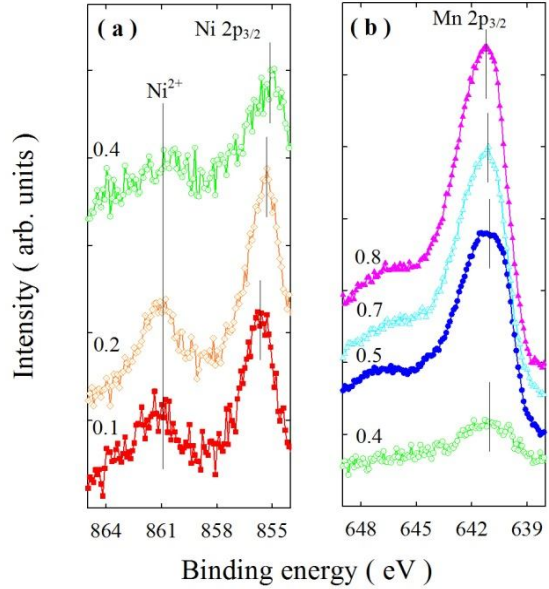

Figure 5. The X-ray photoelectron spectroscopy (XPS) spectra of (a) Ni $2p_{3/2}$ for $CeNi_{0.8-x}Mn_xBi_2$ (x = 0.1, 0.2 and 0.4) and (b) Mn $2p_{3/2}$ for $CeNi_{0.8-x}Mn_xBi_2$ (x = 0.4, 0.5, 0.7 and 0.8). The solid lines indicate the peak positions corresponding to the valence states of Ni and Mn.

In order to evaluate the existing proportion of $Mn^{2+}$ to $Mn^{3+}$ ionic states, the XPS data for x ≥ 0.4 are fitted with Gaussian multi-functions. Because the peak energy for $Mn^{2+}(2p_{3/2})$ and $Mn^{3+}(2p_{3/2})$ are fixed with 640.8 eV and 641.6 eV, respectively, the fitting parameters are the peak intensity A and the peak width w. The fitted curves compared with the experimental data are plotted in figure 6, where the XPS data are well fitted with two Gaussian functions from the $Mn^{2+}$ and $Mn^{3+}$ contributions. The fitted results of A and w are listed in table 1. Even though the A and w parameters are a bit scattered, the ratio of A ($Mn^{3+}/Mn^{2+}$) increases from 2.6 to 14.8, and the ratio of w ($Mn^{3+}/Mn^{2+}$) increases from 1.5 to 3.0 as x increases. This demonstrates that the contribution of $Mn^{3+}$ ions becomes more dominant for higher x. Therefore, we can understand the ferromagnetic ordering via the double-exchange interaction between $Mn^{2+}$ and $Mn^{3+}$ states.

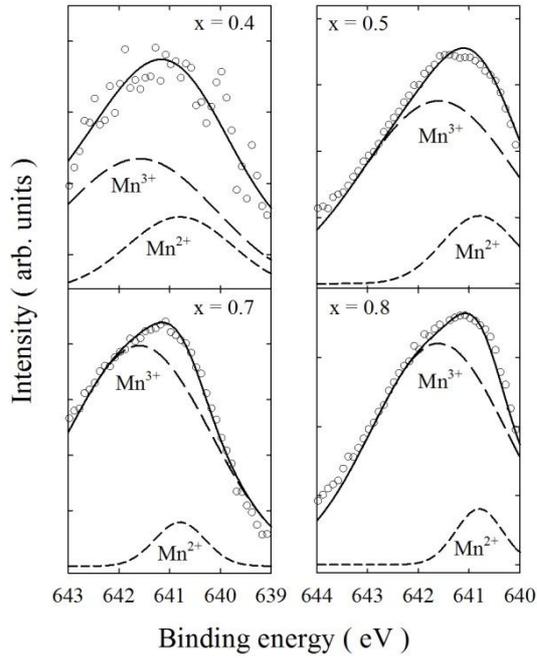

Figure 6. The XPS data (open circles) fitted by two Gaussian functions (dashed lines). The open circles are the experimental data of $CeNi_{0.8-x}Mn_xBi_2$, and the solid lines are the sum of two Gaussian functions where the long and short dashed lines correspond to the curve for $Mn^{3+}$ and $Mn^{2+}$, respectively.

Table 1. The fitting parameters of A and w and the ratio of them, respectively, for x = 0.4, 0.5, 0.7 and 0.8.

| x | Valence state of Mn | A | w | Ratio of A ($Mn^{3+}/Mn^{2+}$) | Ratio of w ($Mn^{3+}/Mn^{2+}$) |
|---|---|---|---|---|---|
| 0.4 | +2 | 2.7 | 2.5 | 2.6 | 1.5 |
| | +3 | 7.0 | 3.6 | | |
| 0.5 | +2 | 9.3 | 1.7 | 5.7 | 2.1 |
| | +3 | 53.0 | 3.6 | | |
| 0.7 | +2 | 4.0 | 1.2 | 14.1 | 2.8 |
| | +3 | 55.8 | 3.3 | | |
| 0.8 | +2 | 5.0 | 1.1 | 14.8 | 3.0 |
| | +3 | 74.1 | 3.5 | | |

Now we discuss our results of inelastic neutron scattering (INS) measurements on CeNi$_{0.8}$Bi$_2$ (i.e. x = 0). First we discuss low energy INS results, which were carried out on OSIRIS to investigate the quasi-elastic response from the CEF ground state of the Ce$^{3+}$ ion. The Ce$^{3+}$ ion having tetragonal point symmetry, CEF potential will split the J = 5/2 ground manifold into three CEF doublets. Thus, low energy INS response at low temperature will give quasi-elastic scattering from the ground state CEF doublet in the paramagnetic state. At high temperature the contribution from higher CEF levels will also contribute to the observed quasi-elastic linewidth.

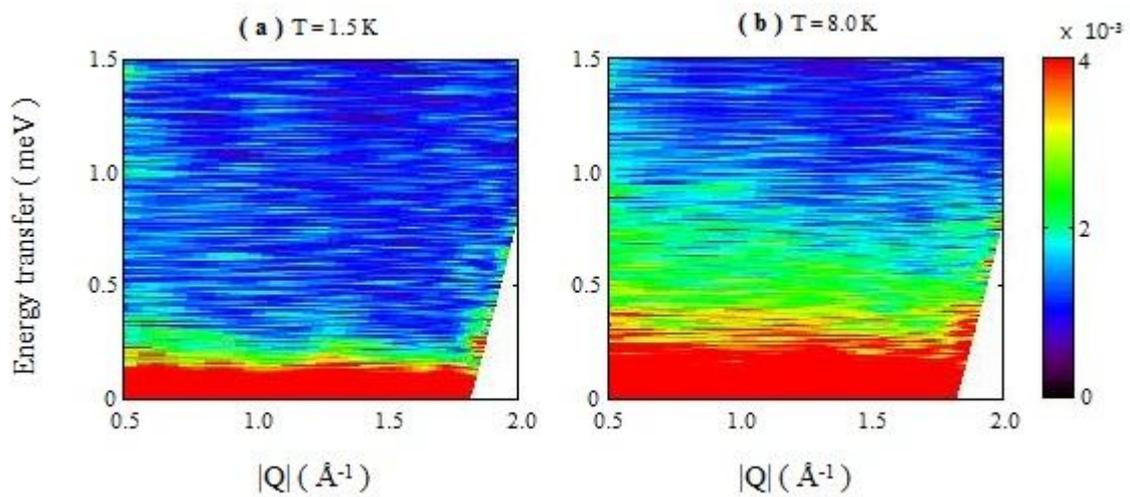

Figure 7. The colour coded contour maps of the scattering intensity at (a) 1.2 K and (b) 8 K from CeNi$_{0.8}$Bi$_2$ measured on OSIRIS using a PG002 analyser with final fixed energy of 1.845 meV.

Figure 7(a) and (b) show colour coded contour maps of the scattering intensity at 1.2 K and 8 K. It is clear from this figure that we have a clear sign of quasi-elastic intensity at 8 K, while the quasi-elastic scattering is reduced or almost disappear at 1.2 K. This reduction in the intensity is associated with the magnetic ordering of the Ce moments, which splits the ground state doublet into two singlets and also the quasi-elastic intensity transfers into the spin wave scattering. Furthermore, we also expect that the observed quasi-elastic/paramagnetic scattering above T$_N$ goes into the elastic magnetic Bragg peaks.

In order to study the temperature dependence of the quasi-elastic linewidth, we have made Q-integrated (|Q| = 1.27 Å$^{-1}$) energy cuts, which are shown in figure 8. To compare the linewidth of elastic peak due to the instrument resolution, we have also plotted the data of vanadium measured in the same conditions (solid line in figure 8) along with the sample data. It shows that at 1.2 K the linewidth of the elastic peak of the sample is resolution-limited, and there is no clear sign of extra contribution from quasi-elastic line. There is a weak sign of inelastic peak centered near 0.7 meV at 1.2 K, which we attribute to powder average spin wave scattering. Further at 3 K and above a clear sign of the quasi-elastic scattering is observed. The quasi-elastic scattering is not changed much between 6 K and 12.5 K (see figure 8).

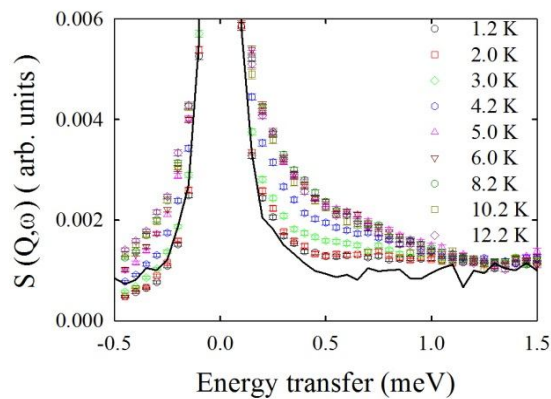

Figure 8. The Q-integrated (|Q| = 1.27 Å$^{-1}$) energy cuts at various temperature from CeNi$_{0.8}$Bi$_2$. We have also plotted the data of a vanadium sample measured with the similar conditions shown by solid line to show the instrument resolution.

Figure 9(a) and (b) show the colour coded intensity maps of inelastic neutron scattering intensity of CeNi$_{0.8}$Bi$_2$ and LaNi$_{0.8}$Bi$_2$ measured at 6 K on MARI. We have also measured both samples at 1.2 K and the data (not shown here) are very similar to that of 6 K. From figure 9, it is clear that at low-Q we have two magnetic excitations centered near 9 meV and 19 meV in CeNi$_{0.8}$Bi$_2$, while at high-Q we have phonon scattering at the same position where we have CEF excitations. This point is very clear in figure 10 where we have plotted Q-integrated energy dependent intensity at low-Q (Q = 1.91Å$^{-1}$) in figure 10(a) and at high-Q (Q = 7.3Å$^{-1}$) in figure 9(b). At low-Q the intensity from LaNi$_{0.8}$Bi$_2$ is smaller than that of CeNi$_{0.8}$Bi$_2$, although the total scattering cross-section of the former is 18% greater than that of the latter. This indicates that at low-Q we have magnetic scattering in CeNi$_{0.8}$Bi$_2$ that arises from the crystal electric field splitting of the J=5/2 state of Ce$^{3+}$ ion under the tetragonal point symmetry. Considering Ce$^{3+}$ having odd number of 4f electron (4f$^1$) Kramers degeneracy theorem, which states that for an odd numbers of electron, the energy levels must be at least doubly degenerate and hence we expect three CEF doublets in the paramagnetic state (i.e. above T$_N$) of CeNi$_{0.8}$Bi$_2$. In order to analyze our data using CEF model, we have estimated the magnetic scattering at 6 K, which is shown in figure 11. The magnetic scattering was estimated by subtracting phonon scattering using LaNi$_{0.8}$Bi$_2$ data as: S(Q,ω)$_M$ = S(Q,ω)$_{Ce}$ - S(Q,ω)$_{La}$ × factor, where the factor (= 0.843) is the ratio of the total scattering cross-section of CeNi$_{0.8}$Bi$_2$ and LaNi$_{0.8}$Bi$_2$. In figure 11, it is clear that we have two well-defined CEF excitations near 9 and 19 meV. Further the linewidth of 19 meV excitation is larger than that of 9 meV excitation, which may be either due to hybridization of this CEF level with that of conduction electrons or coupling between electronic and lattice degree of freedom.

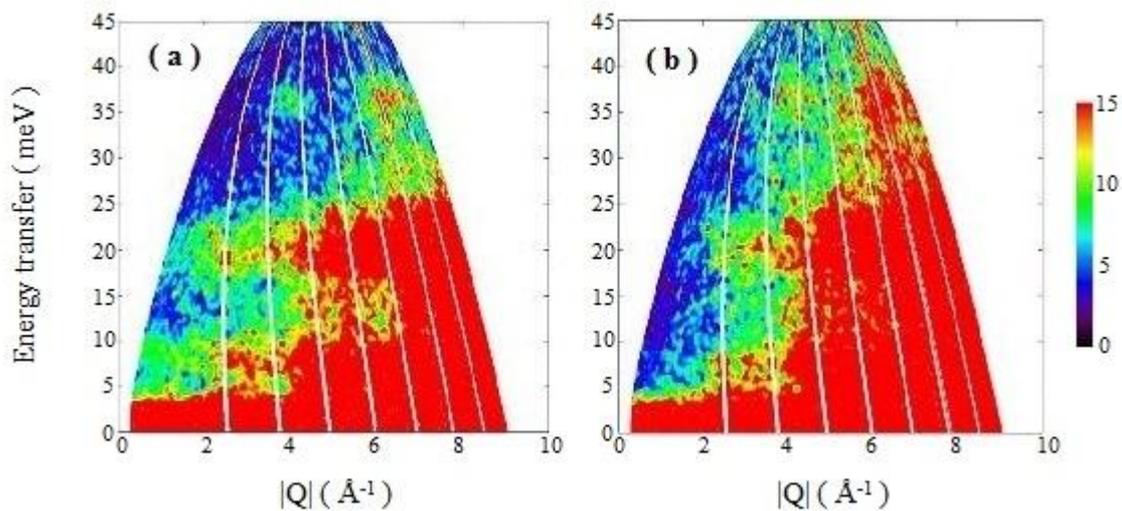

Figure 9. The colour coded intensity maps of inelastic neutron scattering intensity from (a) CeNi$_{0.8}$Bi$_2$

and (b) LaNi$_{0.8}$Bi$_2$ measured at 6 K on MARI using the incident energy of 50meV.

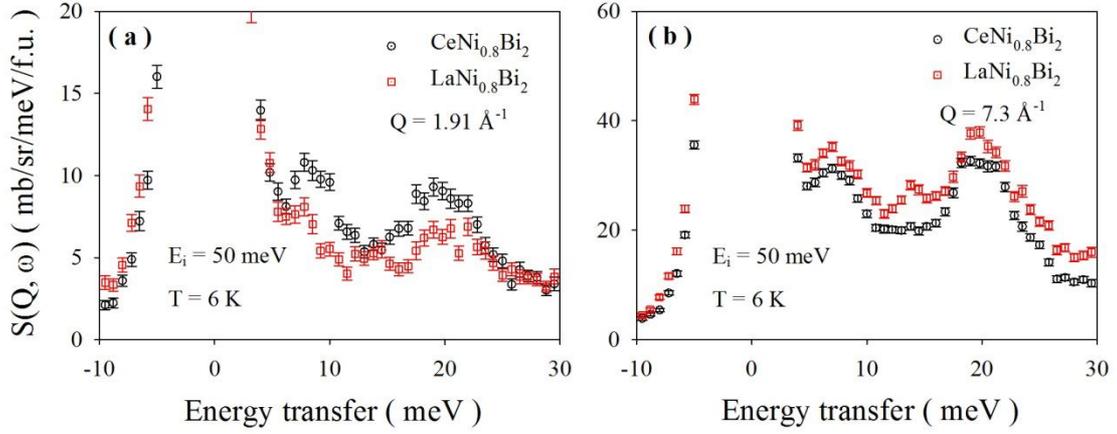

Figure 10. The Q-integrated energy dependent intensity, (a) at low-Q (|Q| = 1.91 Å$^{-1}$) and, (b) at high-Q (|Q| = 7.3 Å$^{-1}$) from CeNi$_{0.8}$Bi$_2$ and LaNi$_{0.8}$Bi$_2$ measured at 6 K on MARI using the incident energy of 50 meV.

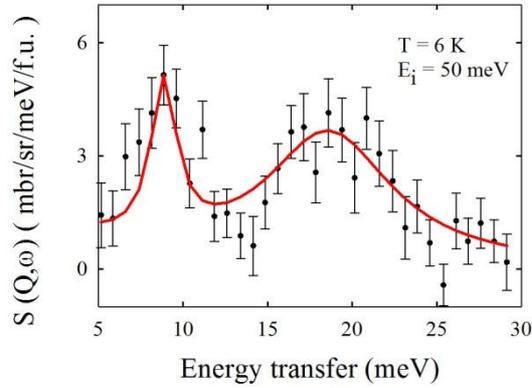

Figure 11. The estimated magnetic scattering from CeNi$_{0.8}$Bi$_2$ at 6 K. The magnetic scattering was estimated by subtracting phonon scattering from LaNi$_{0.8}$Bi$_2$. The solid line shows the fit based on the crystal electric field (CEF) model (see text).

We now proceed with a more detailed analysis of the observed CEF excitations based on the CEF model. In the tetragonal point symmetry, 4mm (C$_{4v}$), at the Ce site, the CEF Hamiltonian can be represented as follows:

$$H_{CF} = B_2^0 O_2^0 + B_4^0 O_4^0 + B_4^4 O_4^4 \qquad (1)$$

where $B_n^m$ are the CEF parameters to be determined from the experimental data, and $O_n^m$ are Stevens operator equivalents obtained using the angular momentum operators [23, 24, 25].

The value of $B_2^0$ can be determined using the high temperature expansion of the magnetic susceptibility [23, 24, 25], which gives $B_2^0$ in terms of the paramagnetic Curie-Weiss temperature, $\theta_{ab}$ for an applied field in the ab-plane and $\theta_c$ for an applied field in the c-axis.

$$B_2^0 = \frac{10}{3} (\theta_{ab} - \theta_c)/(2J-1)(2J+3) \qquad (2)$$

Note that the above equation (2) is valid for isotropic exchange interactions (i.e. isotropic molecular field parameters). Using the values of $\theta_{ab}$ = -156 K and $\theta_c$ = -6 K from the single crystal susceptibility of CeNi$_{0.8}$Bi$_2$ [14], we find $B_2^0$ = -15.63 K (-1.347 meV) from the high temperature Curie-Weiss behavior. In the first stage of our analysis of INS data, we kept a fixed value of $B_2^0$ = -1.347 meV and varied both values of $B_4^0$ and $B_4^4$. However, we could not find good solution that fits well our INS data and single crystal susceptibility data together. This indicates that the magnetic exchanges are anisotropic and hence the $B_2^0$ value estimated from the above equation is not valid. Thereby, in further analysis of INS data we allowed $B_2^0$, $B_4^0$ and $B_4^4$ to vary and fitted the neutron scattering data at 6.2 K and single crystal susceptibility data together. We added anisotropic molecular field parameters ($\lambda^a$ and $\lambda^c$) and temperature independent susceptibility ($\chi_0^a$ and $\chi_0^c$). The values of the parameters obtained from this procedure are $B_2^0$ = -0.3712(3) meV, $B_4^0$ = -0.0273(2) meV, $B_4^4$ = 0.2805(8) meV, $\lambda^a$ = -52.7(12) mole/emu, $\lambda^c$ = -2.1(1) mole/emu, $\chi_0^a$ = 1.3 × 10$^{-4}$ emu/mole, and $\chi_0^c$ = 1.6 × 10$^{-3}$ emu/mole. The fit to the INS data can be seen in figure 11 and the fit to the single crystal susceptibility can be seen in figure 12. The CEF wave functions obtained from the fit are given by

$$\Psi_1^\pm = (-0.452) |\pm \tfrac{3}{2}\rangle + (0.892) |\mp \tfrac{5}{2}\rangle$$

$$\Psi_2^\pm = |\pm \tfrac{1}{2}\rangle$$

$$\Psi^{\pm}_3 = (0.892)|\pm\tfrac{3}{2}\rangle + (0.453)|\mp\tfrac{5}{2}\rangle$$

They correspond to the energy eigenvalues of 0, 8.9 and 18.7 meV. We can now estimate the value of the ground state magnetic moment using the above wave functions and the following formulae:

$$<\mu_x> = <\Psi^{\pm=}_1|\frac{g_J}{2}(J^+ + J^-)|\Psi_1>$$

$$<\mu_z> = <\Psi^{\pm}_1|g_J(J_z)|\Psi^{\pm}_1>$$

The estimated magnetic moments are $<\mu_x> = 0.77\ \mu_B$ and $<\mu_z> = 1.44\ \mu_B$. The value of $<\mu_z>$ estimated here is in good agreement with that obtained from the neutron diffraction study 1.43 $\mu_B$ [13].

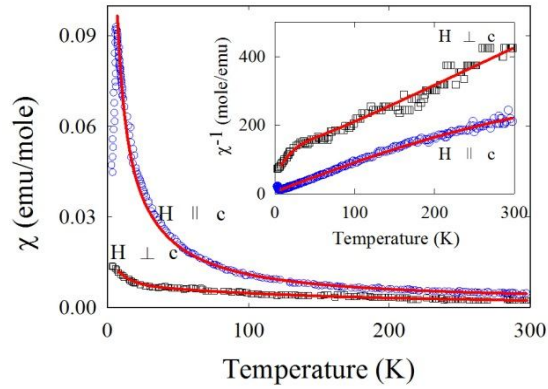

Figure 12. The single crystal susceptibility of CeNi$_{0.8}$Bi$_2$ along a-axis and c-axis from reference [14]. The solid line shows the fit based on the crystal electric field (CEF) model (see text). The inset shows the inverse susceptibility as a function of temperature.

## 4. Conclusion

We have studied the magnetic and transport properties of CeNi$_{0.8-x}$Mn$_x$Bi$_2$. The Mn substitution for Ni brings dramatic changes in their physical properties. With increasing x, the magnetic susceptibility is strongly enhanced, resulting in a ferromagnetic state for x > 0.4 and also, the electrical resistivity is strongly reduced, resulting in a simple metallic state for x > 0.4. These results of ferromagnetism and metallicity can be explained by the double-exchange mechanism from the mixed valence states of Ni and Mn. The X-ray photoelectron spectroscopy experiments reveal that with increasing x (< 0.4) the initial $Ni^{3+}$ state gradually changes to the $Ni^{2+}$ state, which is substituted for the $Mn^{2+}$ state, and for further increase of x (> 0.4) the $Mn^{2+}$ state changes to the final $Mn^{3+}$ state. Furthermore, the inelastic neutron scattering (INS) study reveals two well defined crystal field excitations near 9 and 19 meV, which indicates that Ce ions are in 3+ states with 4f-electrons having localized nature. We have analyzed the INS data and single crystal susceptibility of $CeNi_{0.8}Bi_2$ using a CEF model, which explain the observed direction and the magnitude of the ordered state moment of the Ce ions.

**Acknowledgments**

This work was supported by the Basic Science Research Program through the NRF of Korea funded by the Ministry of Education, Science and Technology (2012R1A1A2039944). D.T.A. acknowledges financial assistance from CMPC-STFC grant number CMPC-09108. We would like to thanks Drs W. Kockelmann and R. Smith for their help in samples characterization on the GEM diffractometer at ISIS Facility.